\documentclass[aps,prb,twocolumn,showpacs,amsmath,amssymb,unsortedaddress]{revtex4-1}
\usepackage{graphicx}
\usepackage{longtable}
\usepackage{ulem}
\usepackage{color}

\begin{document}
\title{Effect of Ion Bombardment on the Chemical Properties of Crystalline Tantalum Pentoxide Films}

\author{Israel Perez} 
\email[Contact Author: ]{cooguion@yahoo.com}
\affiliation{National Council of Science and Technology (CONACYT)-Department of Physics and Mathematics, Institute of Engineering and Technology, Universidad Aut\'onoma de Ciudad Ju\'arez, Av. del Charro 450 Col. Romero Partido, C.P. 32310, Ju\'arez, Chihuahua, M\'exico}
\author{V\'ictor Sosa} 
\affiliation{Applied Physics Department, CINVESTAV Unidad M\'erida, km 6 Ant. Carretera a Progreso, A.P. 73, C.P. 97310 M\'erida, Yucat\'an, M\'exico}
\author{Fidel Gamboa Perera}
\affiliation{{\it Applied Physics Department, CINVESTAV Unidad M\'erida, km 6 Ant. Carretera a Progreso, A.P. 73, C.P. 97310 M\'erida, Yucat\'an, M\'exico}}
\author{Jos\'e Trinidad Elizalde Galindo} 
\affiliation{Institute of Engineering and Technology, Universidad Aut\'onoma de Ciudad Ju\'arez, Av. del Charro 450 Col. Romero Partido, C.P. 32310, Ju\'arez, Chihuahua, M\'exico}
\author{Jos\'e Luis Enr\'iquez-Carrejo} 
\affiliation{Institute of Engineering and Technology, Universidad Aut\'onoma de Ciudad Ju\'arez, Av. del Charro 450 Col. Romero Partido, C.P. 32310, Ju\'arez, Chihuahua, M\'exico}
\author{Pierre Giovani Mani Gonz\'alez} 
\affiliation{Institute of Engineering and Technology, Universidad Aut\'onoma de Ciudad Ju\'arez, Av. del Charro 450 Col. Romero Partido, C.P. 32310, Ju\'arez, Chihuahua, M\'exico}
\date{\today}

\begin{abstract}
The effect of argon ion bombardment on the chemical properties of crystalline Ta$_2$O$_5$ films grown on Si(100) substrates by radio frequency magnetron sputtering was investigated by X-ray photoelectron spectroscopy. All samples were irradiated for several time intervals [(0.5, 3, 6, 9) min] and the Ta $4f$ and O $1s$ core levels were measured each time. Upon analysis at the surface of the films, we observe the Ta $4f$ spectrum characteristic of Ta$_2$O$_5$. Irradiated samples exhibit the formation of Ta suboxides with oxidation states Ta$^{1+}$, Ta$^{2+}$, Ta$^{3+}$, Ta$^{4+}$, and Ta$^{5+}$. Exposing the films, after ion bombardment, to ambient for some days stimulates the amorphous phase of Ta$_2$O$_5$ at the surface suggesting that the suboxides of Ta are unstable. Using a sputtering simulation we discuss that these suboxides are largely generated during ion bombardment that greatly reduces the oxygen to tantalum ratio as the irradiation time increases. The computer simulation indicates that this is due to the high sputtering yield of oxygen.
\end{abstract}
\maketitle
\section{Introduction}
Investigations in high dielectric constant materials point to tantalum pentoxide (Ta$_2$O$_5$) as one of the most promising candidates to deal with several modern technological challenges. Due to its low leakage current and high dielectric constant, tantalum pentoxide has been used in storage capacitors, insulators, catalysts, gas detectors, and memory devices \cite{tkaga91a,kwkwon96,cchaneliere98a,hchen11a}. In addition to its electrical properties, Ta$_2$O$_5$ possesses high refraction index ($n=2.18$ at $\lambda=550 \;\textrm{nm}$) and a wide band gap of $\sim$4.0 eV which makes it an excellent material for optical and electrochromic applications \cite{cchaneliere98b,rhdennard74a,sshibata96a,eatanassova99a,gatak17a,qliu18a,qliu18b,ddong18a}. 

Regarding its atomic structure, Ta$_2$O$_5$ solidifies in either amorphous or crystalline structure; the latter showing two phases below 1500 K \cite{eatanassova99a,jdkruschwitz97a,cchaneliere99a}. The study of the physical and chemical properties of Ta$_2$O$_5$ films is of great importance not only from the technological perspective but also from the scientific one. In previous research the existence of several polymorphs \cite{spgarg96a,ktjacob09a} were established and although most researchers agree that the system crystallizes in either hexagonal ($\delta$ phase) or orthorhombic ($\beta$ phase) structures, the ground-state crystal structure is still under investigation \cite{sperezw16a,yyang18a}. The atomic structure and the properties of Ta$_2$O$_5$ strongly depend on the fabrication methods. In the last two decades experimental techniques such as pulsed laser deposition (PLD), low pressure chemical vapour deposition (LPCVD), direct current (DC) and radio frequency (RF) sputtering, ion assisted deposition (IAD), and electron beam evaporation (EBE) have been used \cite{eatanassova95a,hshinriki91a,skamiyama93a,gqlo93a,ykuo92a,ndonkov11a}. Moreover, to ensure the full oxidation of samples most techniques involve post-deposition heat treatments under oxygen flow and temperatures above 473 K \cite{tdimitrova01a,sjjwu09a,dcristea13a}. If the temperature is high enough, crystalline phases are favoured resulting in a transition of the chemical and physical properties \cite{ndonkov11a,ttsuchiya11a,scwang11a,svjchandra10a}. 

In an earlier work \cite{iperez17a} we investigated the effect of the annealing temperature on the morphological, crystal, vibrational, and optical properties of crystalline Ta$_2$O$_5$ films. Here we wish to extend the study in Ta$_2$O$_5$ films to find out how their chemical properties are modified by ion bombardment. In the past, several researchers have used Ta$_2$O$_5$ films (in some cases a standard BCR-261T) in XPS depth profile experiments to study several aspects of the samples such as film composition, chemical properties, and etch rate for different ion guns, ion energies and/or irradiation fluences \cite{eatanassova98a,vrrmedicherla10a,drbaer10a,nbenito15a,rsimpson17a}. From these studies, it is generally believed that, as the ion bombardment progresses, there is a preferential sputtering of oxygen to be caused by the low atomic number of oxygen that leads to the generation of several oxidation states of tantalum. However, very little is discussed on the processes taking place during irradiation. For instance, no information is known on the magnitude of the sputtering yields of both oxygen and tantalum; i.e., how many oxygen/tantalum atoms are sputtered during irradiation and how much material is displaced to other lattice sites? Also, it is not well understood how the atomic lattice is modified during ion irradiation and in turn how this leads to the generation of the different oxidation states of tantalum. In this sense, the aim of this research is to shed some light on these issues that so far have not been completely elucidated.

For this purpose we use an ion argon gun (with a fixed acceleration energy) from an XPS system, and sputter the sample surface of some crystalline Ta$_2$O$_5$ films for several time intervals. Keeping in mind that argon is inert we then assess the effect of ion bombardment on the sample surface and obtain in situ the changes in the chemical states of Ta. The advantage of using the ion gun of the XPS system is that, samples remain in the same chamber under ultra high vacuum ($\sim 10^{-9}$ mbar) conditions, avoiding in this way any contamination or any undesired oxidation process that otherwise would arise when using an external ion gun. Furthermore, the atomic processes involved in ion bombardment at the film surface are analyzed with the aid of a computer simulation that throws very important quantitative and qualitative information. The simulation turns out to be very useful not only to build a wider picture of such processes but also to enrich the interpretation of XPS measurements when depth profiles are performed. 

To achieve our goal, three Ta films were grown on Si(100) substrates by RF magnetron sputtering and the crystalline phase, i.e., Ta$_2$O$_5$, was induced by exposing the films to post-deposition heat treatments at 1273 K in ambient atmosphere. The crystalline structure and microstructure were determined by X-ray diffraction (XRD) and scanning electron microscopy (SEM), respectively. XPS in combination with an ion gun were used to evaluate the chemical properties for different sputtering time intervals. Finally, using the actual data from the ion irradiation we carried out the simulation and discuss the bombardment process in detail.

\section{Experimental}
\subsection{Film growth and annealing}
Three amorphous Ta films were grown at different sessions at room temperature on 5 mm $\times$ 7 mm--Si(100) substrates by the RF magnetron sputtering technique. Before growth, the substrates were washed with baths in ethanol, acetone, and distilled water. Then each substrate was placed on a rotary base inside the vacuum chamber at a target-to-substrate distance of 12 cm. For the deposition we used a 2.5 inch-Ta target (99.95\% purity) and evacuated the vacuum chamber to a base pressure of $6.6 \times 10^{-5}$ mbar. To activate the plasma, argon gas was flushed into the chamber to a partial pressure of ($2.6\pm 0.1)\times10^{-2}$ mbar. The native oxide layer on the target surface was removed by a 5 min pre-sputtering process at 60 W. Immediately the power was raised to 120 W and the rotary base was activated with a speed of 0.2 rpm. During deposition there was no intentional substrate heating or cooling and no oxygen was injected into the chamber. To oxidize the samples, two films were exposed to post-deposition heat treatments in air for one hour at 1273 K using a Thermo Scientific Thermolyne cylindrical furnace (model F21135). These films were labeled as F05 and F33. The as-deposited film, labeled F0, was kept for reference and was not exposed to any heat treatment. The deposition rate was determined using the deposition time and the sample thickness which was previously measured in the SEM. The film thicknesses were (2.4, 0.5 and 3.3) $\mu$m for F0, F05, and F33, respectively. The deposition rates were 2.8 \AA/s for F0 and F33, and 2.4 \AA/s for F05.

\subsection{Crystal structure and morphology}
After exposing the samples to heat treatments, their crystalline structure was studied by a Siemens diffractometer model D-5000 with Cu $K_{\alpha}$ radiation ($\lambda=1.5406$ \AA). A Bragg-Bretano configuration was used and XRD patterns were obtained at steps of 0.02$^\circ$ with a time per step of 3 s and operating parameters of 34 kV and 25 $\mu$A. A field emission microscope JSM7000F was used to evaluate the morphology and grain size of the samples. 

\subsection{Ion bombardment and XPS characterization}
To study the effect of ion bombardment on the chemical properties of Ta$_2$O$_5$ films, the samples were placed inside the vacuum chamber of a Thermo Scientific K-Alpha XPS spectrometer at a base pressure of $5\times10^{-9}$ mbar. As a sputtering agent we used argon with a purity of 99.998\%. Argon ions were accelerated with a voltage of 3 kV by an ion gun EX06 Thermo Scientific, generating an electric current of 10 $\mu$A. The incidence angle between the sample and the ion gun was 90$^\circ$. The etched area was 2 mm$^2$ and the samples were sputtered at different time intervals ($t_{sput}$), namely: (0.5, 3.0, 6.0,  9.0) min. To evaluate the effect of the ion bombardment on the film surface, simulations were carried out with the SRIM/TRIM software \cite{jpbiersack80a,jfziegler89a}. This software gives, in real time, information on the total and partial sputtering yield, damage area, number of sputtered atoms, among other magnitudes. Particular settings for the simulations are given in the corresponding section below.

After each sputtering session, the spectra of the Ta $4f$ and O $1s$ core levels were measured in the same instrument using an Al K$_{\alpha}$ X-ray source with photon energy of 1486.7 eV, and set to 12 kV and 40 W. The X-ray beam spot has a diameter of 400 $\mu$m and makes an angle relative to the sample of 30$^\circ$. For the XPS scans we used steps of 0.1 eV. Both the stoichiometry of the films and the oxidation states of Ta were determined by peak deconvolution of the XPS spectra using Voigt functions as implemented in the AAnalizer software (see below for more details) \cite{aherrerag14a}.

\section{Results and discussion}
\subsection{Crystalline structure}
\begin{figure}[t!]
	\begin{center}
		\includegraphics[width=10cm]{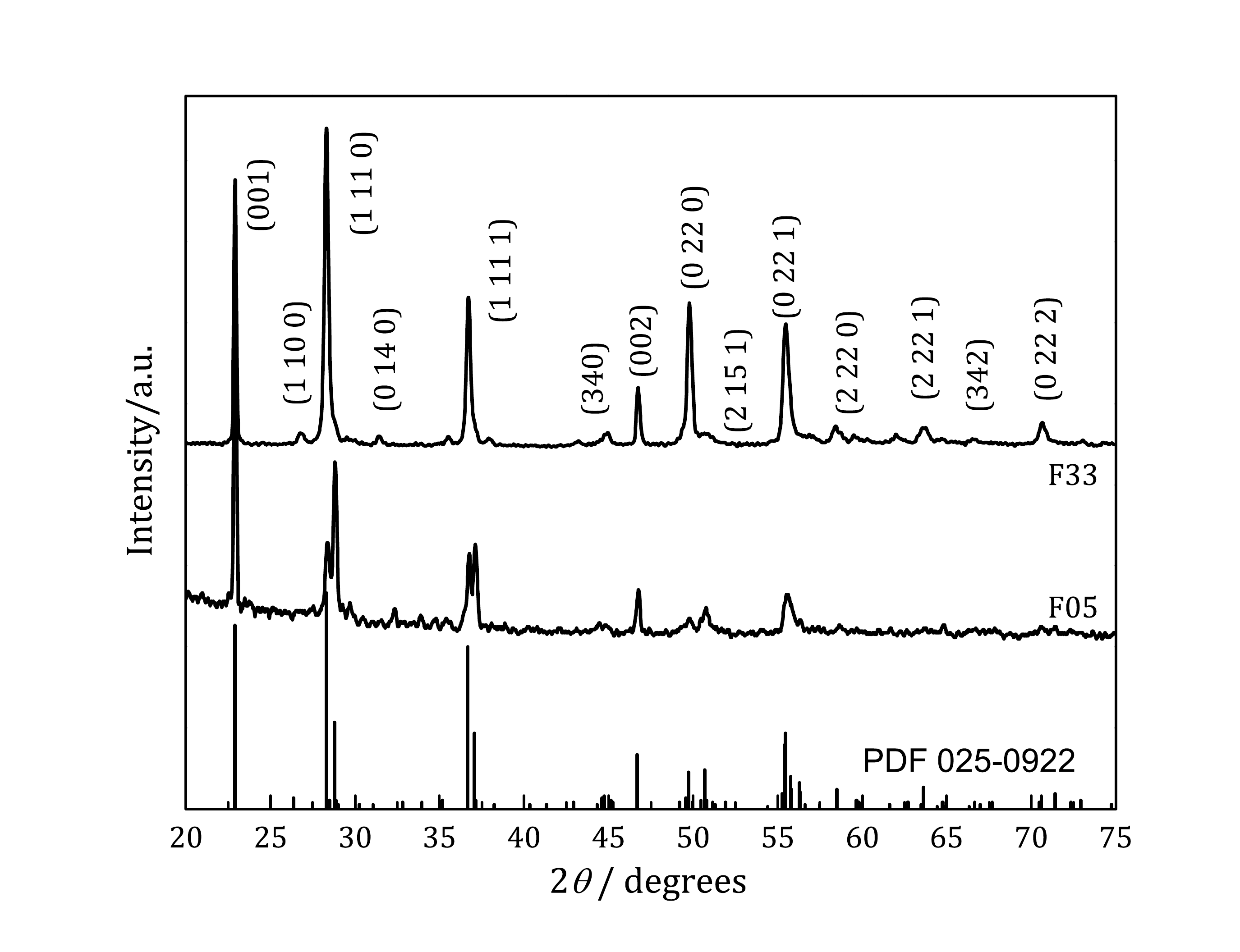}
		\caption{X-ray diffraction patterns for the crystalline films, F05 and F33. The lower pattern corresponds to the reference PDF 025-0922}
		\label{fig1}
	\end{center}
\end{figure}

As expected, the as-deposited film F0 displays no peaks in the XRD pattern indicating an amorphous structure; for this reason its pattern is not shown here. On the contrary, the diffraction patterns for F05 and F33 imply a crystalline structure (see Figure \ref{fig1}).
These patterns can be indexed to the orthorhombic phase $\beta-$Ta$_2$O$_5$ (PDF 00-025-0922 with lattice parameters $a=6.1980$ \AA, $b=40.2900$ \AA, $c=3.8880$ \AA, and $\alpha=\beta=\gamma=90^\circ$ for the spatial group $P2_12_12$). We stress that there is still an ongoing discussion regarding the specific space group for the crystalline phases of Ta$_2$O$_5$  \cite{sperezw16a,yyang18a,iperez17a,jykim14a,shlee13a,jlee14a,zhelali14a,yguo15a,jykim15a}.

It is worth noting, however, that the patterns show no traces of other Ta oxides. According to the literature \cite{spgarg96a}, Ta$_2$O$_5$ is the most stable oxide of Ta exhibiting amorphous and crystalline phases; the other oxides such as TaO, Ta$_2$O, TaO$_2$, Ta$_2$O$_3$ (except for TaO$_{x}$) have shown to be difficult to obtain as pure phases. Moreover, except for Ta$_2$O$_3$ and TaO$_x$, all suboxides are crystalline. We will come back to this in our discussion on chemical states.

The crystallite size $D$ for the samples F05 and F33 was estimated using the celebrated Scherrer equation
\begin{equation}
\label{sch}
D=\frac{K\lambda}{\Gamma \cos\theta},
\end{equation}
where $K=0.9$, $\lambda=1.5406\;\textrm{\AA}$ and $\Gamma$ is the full width at half maximum of a given peak. Accordingly, taking the peak at $2\theta=28.3^\circ$ for both films, we found that $\Gamma$=(0.22, 0.27)$^\circ$ and hence $D=$(36, 30) nm, respectively. 

\subsection{Morphology and grain size}
\begin{figure}[t!]
	\begin{center}
		\includegraphics[width=5.4cm]{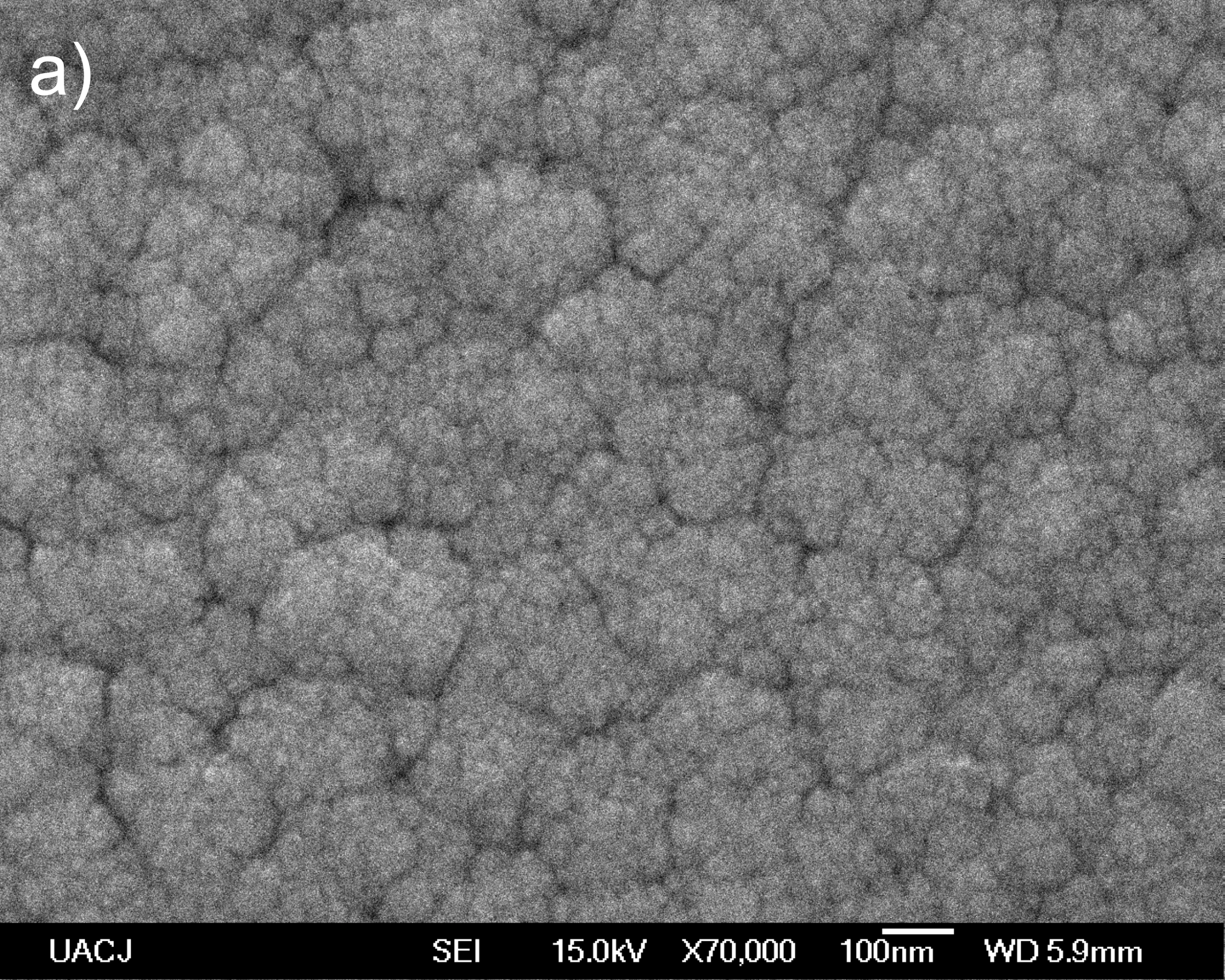} \includegraphics[width=5.4cm]{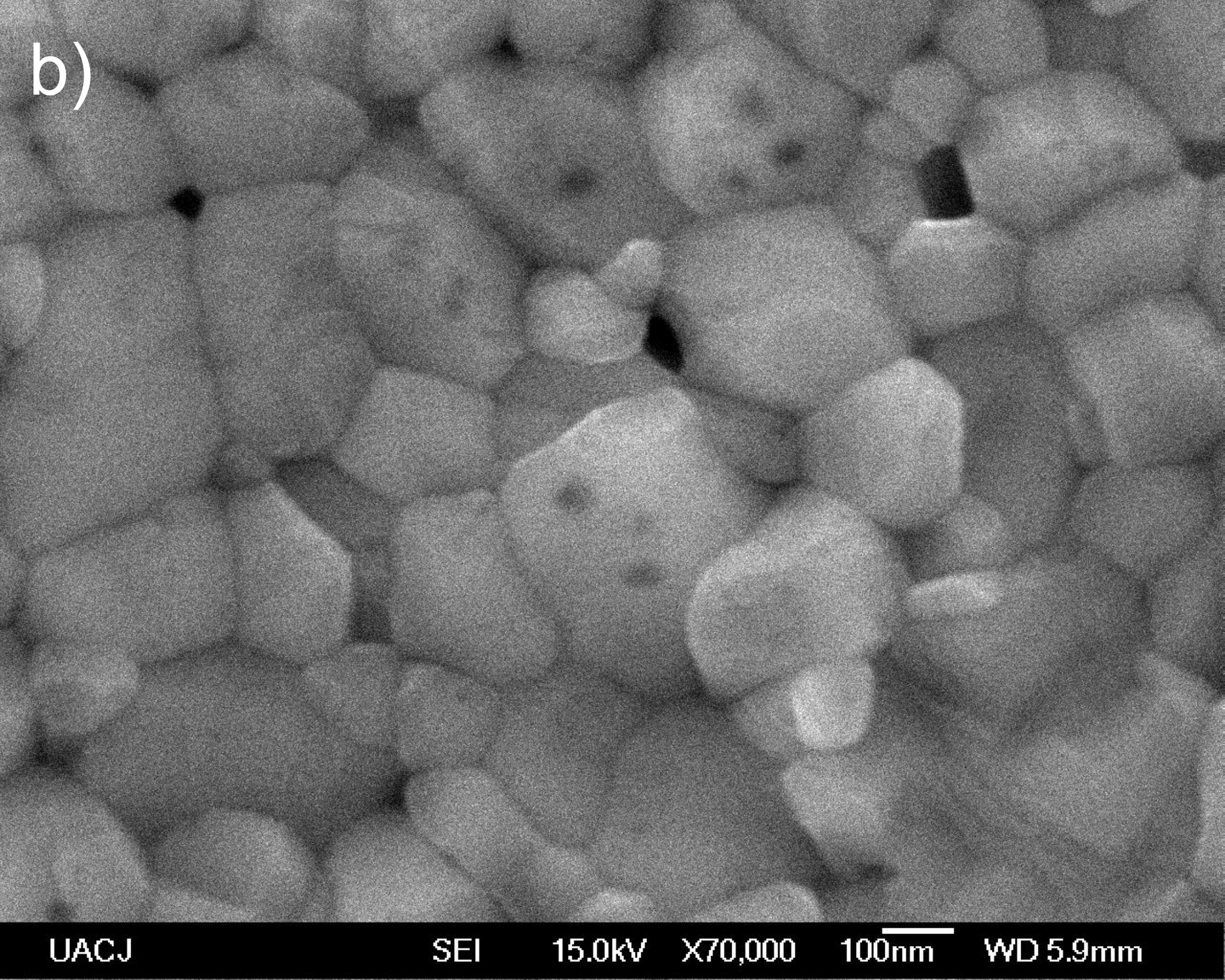} \includegraphics[width=5.4cm]{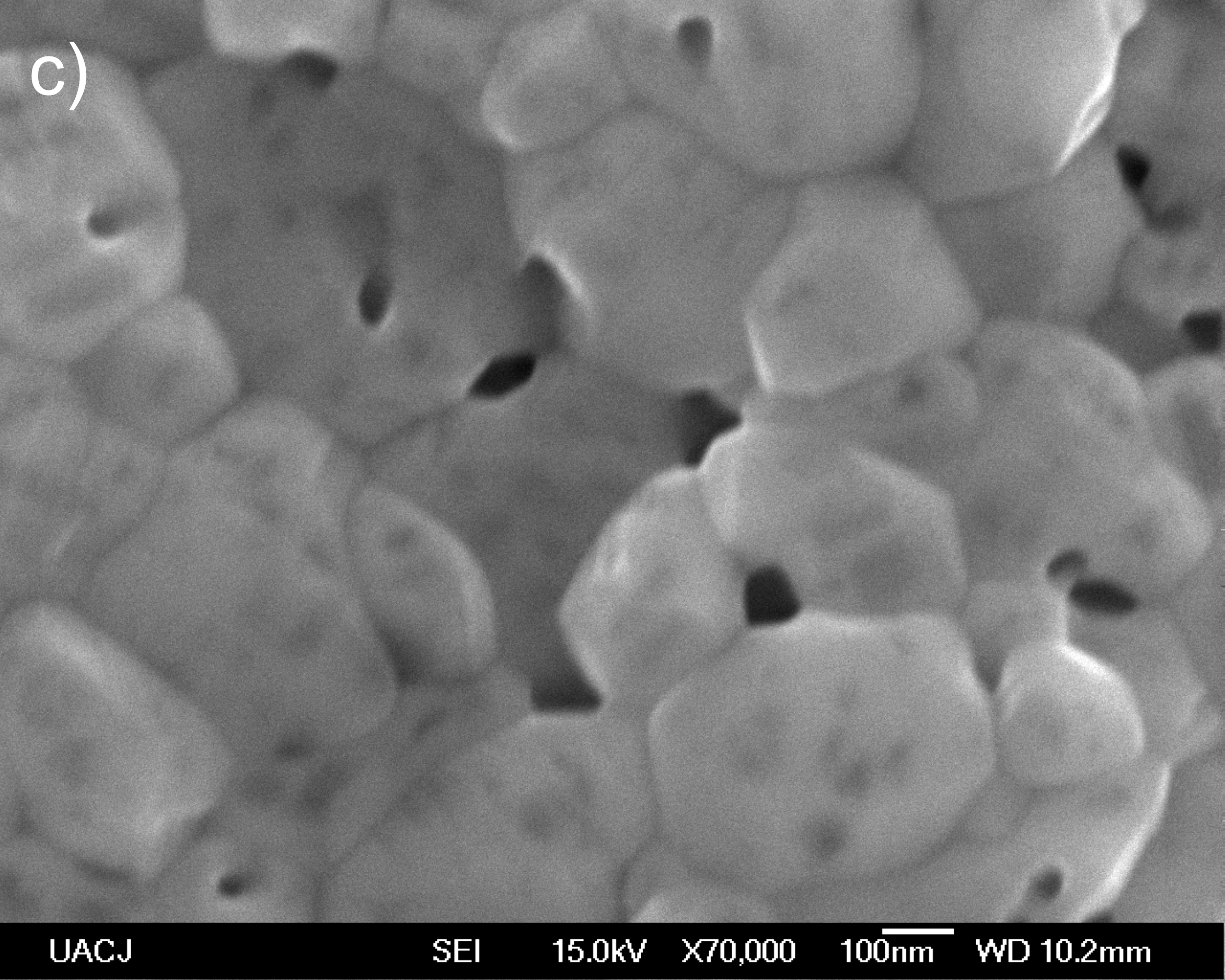}
		\caption{SEM images of the films: F0 (a), F05 (b), and F33 (c). F0 exhibits a cloudy-like microstructure whereas the others show randomly oriented grains with black spots}
		\label{S17Ta2O5}
	\end{center}
\end{figure}

In Figure \ref{S17Ta2O5} the SEM images for the three films are displayed. F05 and F33 are shown after annealing. F0 (Figure 2a) shows a smooth and cloudy-like pattern and this same pattern was exhibited by the films F05 (Figure 2b) and F33 (Figure 2c) before annealing. As the films were exposed to the heat treatment, large grains started to show up, both samples resembling the powder microstructure of Ta$_2$O$_5$. We note that both samples exhibit some black spots. 
To elucidate the nature of these spots we took backscattering electron images (not shown). The results suggest that the spots may belong to another oxide phase of Ta, most probably the $\delta$ phase of Ta$_2$O$_5$ or a suboxide. We performed an EDS (Energy dispersive x-ray spectroscopy) analysis on these spots but the atomic concentration of O and Ta did not change significantly and therefore conclusions could not be drawn.  As for the grain size (not to be confused with the crystallite size above), according to these images, we estimated a diameter varying from 100 nm to 500 nm with a mode of 300 nm for F05 and F33 which is a relatively large size for a grain and this allows us to consider our films as granular bulk materials. Despite that both films have different thickness, the grain size and morphology are quite similar as well as the manifestation of the black spots, indicating that the fabrication process is quite reproducible.

\subsection{Chemical states by ion bombardment}

\subsubsection{Ta $4f$ core level}
To investigate the chemical states of the films before and after ion bombardment, we carried out XPS measurements on all samples and deconvoluted the spectra for the Ta $4f$ and O $1s$ core levels. The binding energies of our spectra were calibrated with respect to the O $1s$ peak at 532 eV. After calibration we checked for surface contamination from the C $1s$ core level. 
\begin{figure}[b!]
	\begin{center}
		\includegraphics[width=10cm]{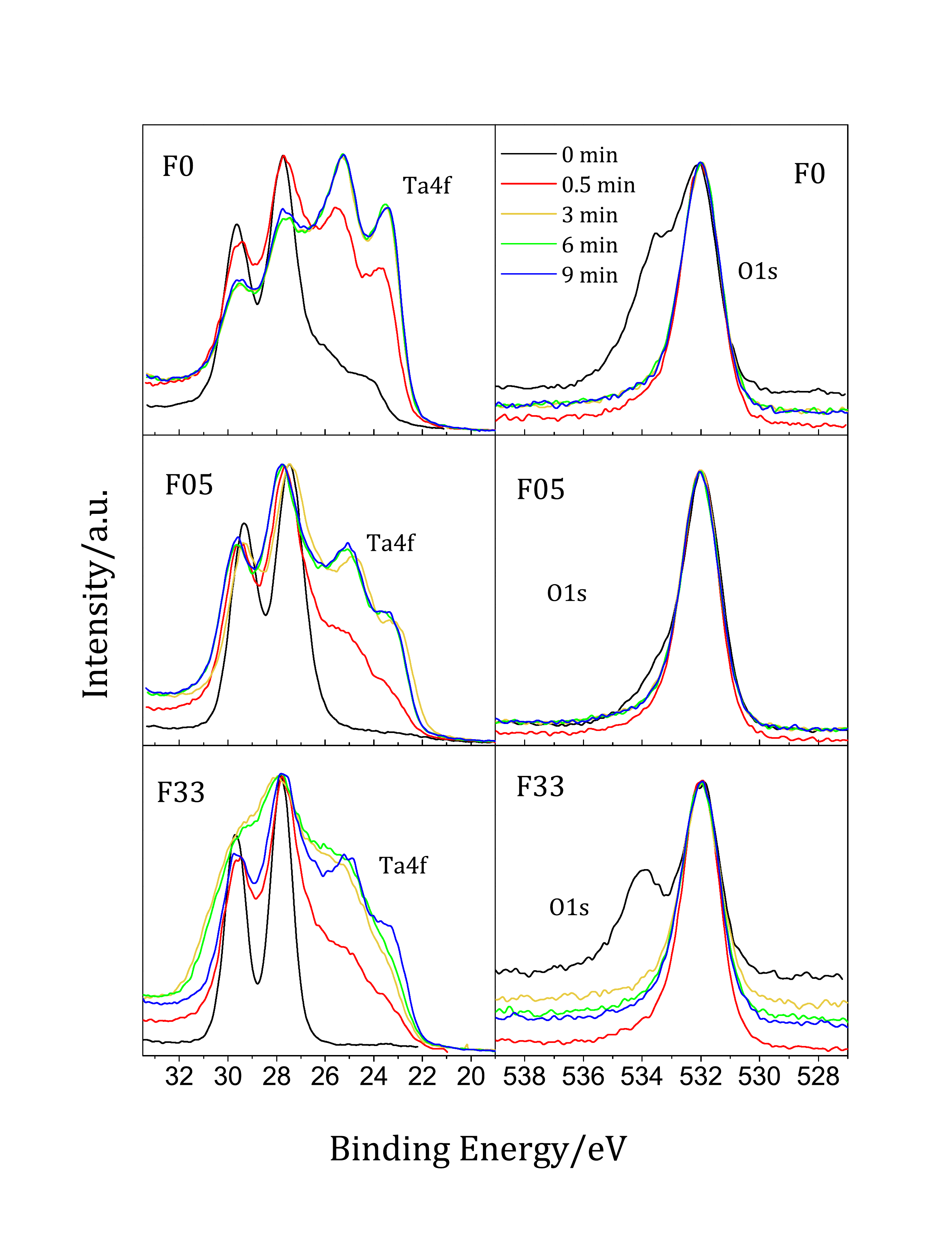} 
		\caption{Normalized to height XPS spectra for Ta $4f$ (left) and O $1s$ (right) for the three films}
		\label{ta4fo1s}
	\end{center}
\end{figure}
The carbon spectra (not shown) revealed the presence of carbon contamination which is quite ubiquitous in most samples (adsorbed carbon). We stress that we have used the O $1s$ binding energy as reference because oxygen, in our samples, has just two contributions (see below). The first one can be associated to O-Ta bonding and the second one to O-C bonding; and the O $1s$ core level is well known to appear at 532 eV. This energy is a good candidate for reference because we are interested in analyzing variations in the Ta $4f$ binding energy and they can be detected with respect to our set of samples, and due to native carbon, the C 1s binding energy would not be a good choice for reference.

The spectra for Ta $4f$ core level as a function of sputtering time are given in the left column of Figure \ref{ta4fo1s}. Starting with the Ta $4f$ core level for the three films without sputtering (0 min, black spectra) we observe that there is a spin-orbit doublet corresponding to the levels $4f_{7/2}$ and $4f_{5/2}$. The binding energy for the doublets is around (27.8$\pm$0.1) eV with spin-orbit splitting of (1.9$\pm$0.1) eV in agreement with those reported for Ta$^{5+}$ in stoichiometric amorphous Ta$_2$O$_5$ films \cite{rsimpson17a,kchen97a,amuto94a}, suggesting that at the surface Ta$_2$O$_5$ forms. This is expected because during annealing the film surface is in an oxygen rich environment that favours surface oxidation much better than in deeper layers. We checked this by sputtering the samples surface for 30 s and observed the red spectra where new features appear. We then exposed the samples to ambient for seven days and observed again the same black spectra without sputtering; an indicative that in the surface the oxidation state +5 is recovered. A closer look at the spectrum for zero etching of F0 and the spectra of F05 and F33 for 30 s etching shows the presence of a shoulder in the low energy region (around $24.6\;\textrm{eV}\pm 0.3$ eV). This signal is known to be caused by screening of $5d$ electrons in amorphous TaO$_x$ ($x=1.86,2.00$) suggesting the presence of TaO$_x$ in our films \cite{ttsuchiya11a}.

\subsubsection{Ion bombardment analysis}
Before going further, it is useful to have a wide picture of the ion bombardment process as a function of sputtering time. This will help us to acquire a deeper understanding of the atomic configuration involved and how it affects the chemical states of the samples. For this goal we carried out a sputtering simulation using the transport of ions in matter (TRIM) package in the surface sputtering/monolayer collision steps mode \cite{jpbiersack80a,jfziegler89a}. This mode gives a full treatment of sputtering for full damage cascades based on Monte Carlo methods. The simulation assumes a stoichiometric target of Ta$_2$O$_5$ with a thickness of 100 nm and bulk density of 8.2 g/cm$^3$. Although our films are thicker than 100 nm, this thickness was selected to make calculations much faster since simulations for thicker targets give the same results but in longer times. During the simulation the film is bombarded by argon ions, impinging normal to the target surface. The ion energy was set to the actual energy of 3 keV; and the ion direction is assumed to be parallel to the $X$-axis with the target surface parallel to the $YZ$ plane, so the penetration depth ($P$) is measured along the $-X$ axis.
\begin{table}[b!]
	\centering 
	\caption{Sputtering parameters for the simulation: Sputtering time ($t_{sput}$), number of incident ions $N$ in a nm$^2$, damage area ($A$), penetration depth ($P$), sputtering yields ($Y$) for Ta and O, and number of sputtered atoms ($N_{sput}$)}
	\label{table2}
	\begin{tabular}{ccccccc}
		\hline
		% after \\ : \hline or \cline{col1-col2} \cline{col3-col4} ...
		$\frac{t_{sput}}{\text{min}}$   &$N$ & $\frac{A}{\text{nm}^2}$  & $\frac{P}{\text{nm}}$ & $Y_{\text{Ta}}$ & $Y_{\text{O}}$ & $N_{sput}$\\ \hline \hline
		0.5 & 960 &85 & 12 & 0.64 & 4.32 & 4800 \\
		3 & 5760 &139 & 13 & 0.63 & 4.37 & 28 800 \\
		6 & 11 520 &165 & 14  & 0.64 & 4.33 & 57 600 \\
		9 & 17 280 & 180 & 14.7  & 0.65 & 4.34 & 86 400 \\
		\hline
	\end{tabular}
\end{table}
\begin{figure*}[t!]
	\begin{center}
		\includegraphics[width=6.5cm]{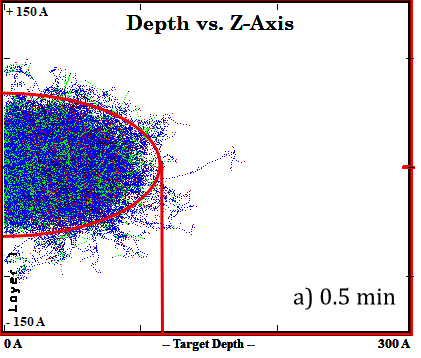} \hspace{1cm} \includegraphics[width=6.5cm]{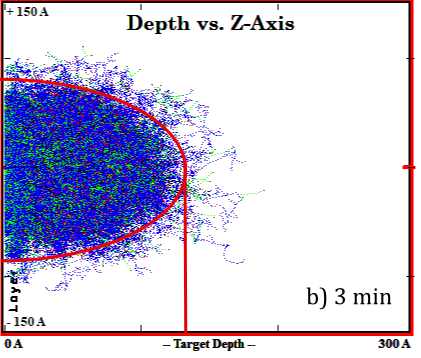} \includegraphics[width=6.5cm]{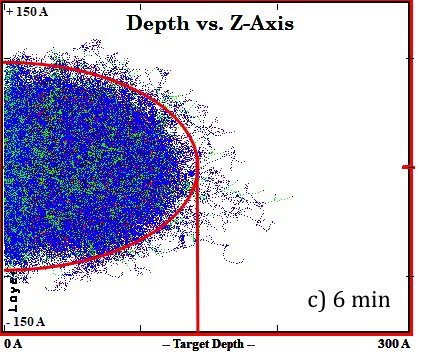} \hspace{1cm} \includegraphics[width=6.5cm]{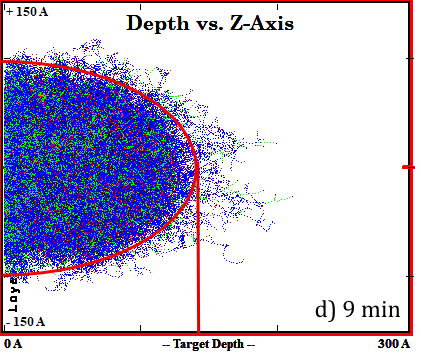}
		\caption{Simulation of damage areas for the different sputtering times along the XZ plane: a) 0.5 min, b) 3 min, c) 6 min, and d) 9 min. Blue dots represent recoiling oxygen atoms and green dots represent recoiling tantalum atoms. Red dots represent stopped Ta and O atoms. The red vertical line is a projection to the horizontal used to determine the penetration depth and the half ellipsis are used to delimit an average damage area}
		\label{damage}
	\end{center}
\end{figure*}
The number of incident ions ($N$) was computed using the actual ion gun current of 10 $\mu$A. Accordingly, we determined that for a surface of 1 nm$^2$, about 32 ions arrive per second. We just then multiplied this number by the corresponding sputtering time intervals, that is: $t_{sput}$=(0.5, 3, 6, 9) min. Table \ref{table2} gives the parameters in each case. 

During the bombardment several processes occur. When an ion arrives at the surface, surface atoms may be either knocked out of the material (sputtered) or displaced deeper into the lattice (implanted). When an ion has a head-on collision with a surface or lattice atom, very often the collision is elastic, otherwise is inelastic and the ion keeps moving and interacting with other lattice atoms. The recoiling atoms (atoms displaced after a collision with either ions or displaced atoms), in turn collide with other lattice atoms, losing energy as they propagate and causing an atomic cascade. This cascade is depicted in Figure \ref{damage} (a-d) in blue dots for O atoms and in green dots for Ta atoms for the different sputtering times (or equivalently, for different $N$). It is evident that most recoiling atoms are O atoms. As the recoiling atoms collide they lose energy and finally stop at some place within the lattice (red dots). When atoms finally stop they may reside either at an interstitial site or at any other lattice site different from the initial one. In this sense, it is likely that a recoiling atom may replace another lattice atom of the same species. If the impinging atom has an energy less than either the lattice binding energy ($E_{\text{bin}}$) or the surface binding energy ($E_{\text{sbe}}$), the atom gives off energy to the lattice/surface atom forcing it to vibrate and release the excess energy as phonons. In our case we use $E_{\text{bin}}=3$ eV for both Ta and O and $E_{\text{sbe}}$ is 2 eV for O and 8.1 eV for Ta. So a 3 keV ion would easily displace target atoms.

The zone affected by the ion bombardment where most atoms recoil is called the damage zone. The images (a-d) show such zone along the $XZ$ plane. The effects along the $XY$ plane are similar and, on average, they are the same as those on the $XZ$ plane and for this reason they are not shown here. In a three dimensional view, the damage zone would resemble a half ellipsoid (the damage ellipsoid). The red semi-ellipse delimits an {\it average} damage area $A$ defined for quantification purposes. This area is a measured of the displaced material during ion bombardment. As summarized in Table \ref{table2}, this quantity increases as $t_{sput}$ increases. Roughly speaking, the damage area doubles its size from 0.5 min to 9 min. The penetration depth in turn (delimited by the red vertical line), varies little for it increases about 1 nm from 0.5 min to 3 min and just 1.7 nm from 3 min to 9 min. Simulations realized by us (not shown) with higher ion energies indicate that the penetration depth largely depends, as expected, on the ion energy and, as shown here, little on the sputtering time. The simulation also gives the sputtering yield ($Y$) per incident ion; the values are 0.64 for Ta and 4.34 for O. For this reason, the number of atoms that leave the sample surface ($N_{sput}$) runs from 4800 for 30 s to 86 400 for 9 min; and so roughly 88\% of the sputtered atoms are oxygen atoms and the rest are Ta atoms. According to our simulation the preferential sputtering of oxygen is not only a consequence of the low atomic number of oxygen, as generally believed \cite{vrrmedicherla10a,drbaer10a,nbenito15a,rsimpson17a}, but also a result of the low surface binding energy of oxygen which causes oxygen atoms to be more prone, than tantalum atoms, to undergo either displacement or sputtering. Therefore, one would expect the depletion of O atoms and both the temporary generation of O vacancies on the surface and the formation of Frenkel defects in the lattice.

With this information in mind, we can figure out the possible effect on the chemical properties. As we can see the effect of sputtering does not extend more than 20 nm deep; so what happen to the chemical environment of the damage zone after the ion bombardment stops? Unfortunately, the simulation does not tell us anything on this matter. It just informs us that many atoms are sputtered from the grains and many others are displaced to other locations within the lattice, leaving a great amount of structural defects and in particular low coordination numbers. So, this scenario strongly suggests that the atomic structure in that zone is highly and energetically unstable. 

To recover stability, several processes must occur to maintain not only the structural stability at the grains surface but also electroneutrality. The atoms in the damage zone must seek for the most favourable energetic sites and, due to oxygen depletion and the large amount of punctual defects, one would expect, in first instance, new coordination geometries leading to the transition of Ta from Ta$^{5+}$ to other oxidation states; and, consequently, the formation of metastable phases of Ta oxides such as TaO, Ta$_2$O, TaO$_2$, TaO$_{x}$, or Ta$_2$O$_3$. 
\begin{figure*}[t!]
	\begin{center}
		\includegraphics[width=17.4cm]{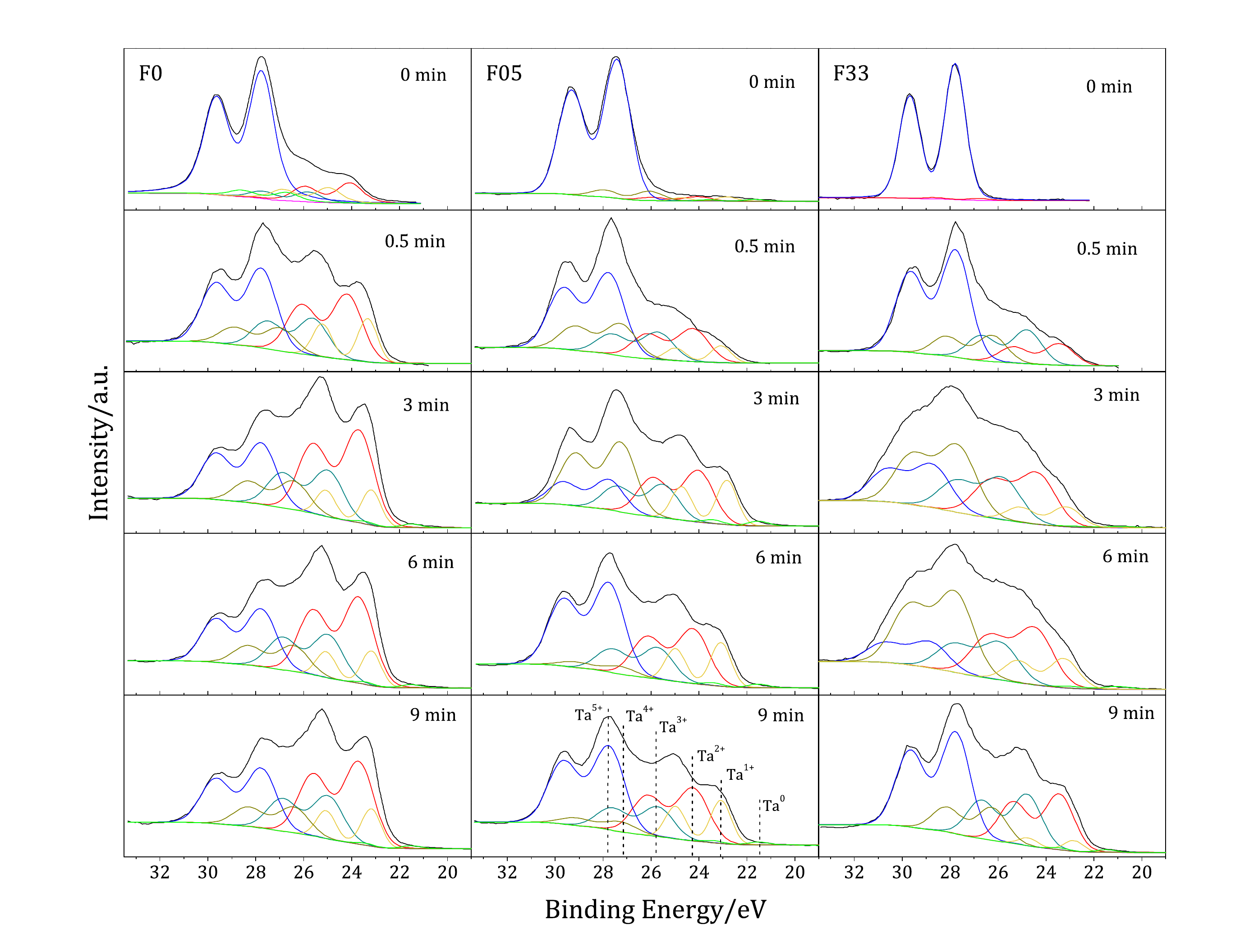} 
		\caption{Deconvolution of Ta $4f$ XPS spectra for the three films. Left column for F0, central column for F05, and right column for F33}
		\label{deconvta4f}
	\end{center}
\end{figure*}
Later, as time goes by (it could be some minutes or days depending on both the oxygen concentration around the surface and the physi- and chemisorption processes taking place \cite{ttsuchiya11a}), the surface tends to readsorb O atoms from the environment and, as we have just discussed above, the surface tends to oxidize in the most stable phase of Ta oxide, i.e., Ta$_2$O$_5$. In this scenario one realizes that after sputtering, there exists amorphous Ta$_2$O$_5$ at the very surface of the grains; and for the rest of the bulk, there is mainly TaO$_x$ for F0 and the $\beta$-Ta$_2$O$_5$ phase for F05 and F33. Hence, as one sputters away the topmost atoms, one destroys the crystalline Ta$_2$O$_5$ phase and the chemical states of Ta change for some time until amorphous Ta$_2$O$_5$ is recovered \cite{eatanassova03a}. The rich additional features that appear in the spectra as the sputtering time increases lends support for this line of thought (0.5 min to 9 min spectra in Figure \ref{ta4fo1s}). 

\subsubsection{Deconvolution analysis of Ta $4f$ core level}
To shed some light on such features, we conducted an analysis of the Ta $4f$ spectra with the software AAnalizer. Because of spin-orbit coupling, the deconvolution process was done with doublets. The latter were fitted using Voigt functions and a Shirley-Sherwood background \cite{aherrerag14a}. The deconvolution analysis reveals the existence of six sets of peak splitting (see Figure \ref{deconvta4f}) indicating the presence of the six oxidations states of Ta, from Ta$^0$ to Ta$^{5+}$. The fitting parameters for the doublets were the following: Gaussian=1.43 eV for the oxidation states +2, +3, +4, +5 and Gaussian=0.9 eV for the oxidation states 0 and +1; and Lorentzian=0.02 eV for all doublets. This kind of features has been reported in ab-initio calculations for the $\delta$ phase of Ta$_2$O$_5$ films \cite{eatanassova98a,ivanov11a,ivanov11b} and have been described by four doublets with the $4f_{7/2}$ binding energies of (22.0, 23.2, 24.6, 26.1) eV and spin-orbit splitting of 1.9 eV. They were attributed to Ta$^{1+}$, Ta$^{2+}$, Ta$^{3+}$/Ta$^{4+}$, and Ta$^{5+}$, respectively, observed in amorphous Ta$_2$O$_5$ films. In our case the six doublets can be associated to the states Ta$^{0}$, Ta$^{1+}$, Ta$^{2+}$, Ta$^{3+}$, Ta$^{4+}$, and Ta$^{5+}$ with binding energies at (21.7$\pm$0.3, 23.0$\pm$0.2, 23.8$\pm$0.4, 25.2$\pm$0.4, 26.2$\pm$0.2, 27.8$\pm$0.2) eV, respectively, and peak splitting of (1.9$\pm$0.1) eV each. The peak positions are slightly higher than those reported earlier but follow the same trend. These discrepancies may be due to differences in the reference spectra used for each measurement.

\begin{figure}[t!]
	\begin{center}
		\includegraphics[width=9cm]{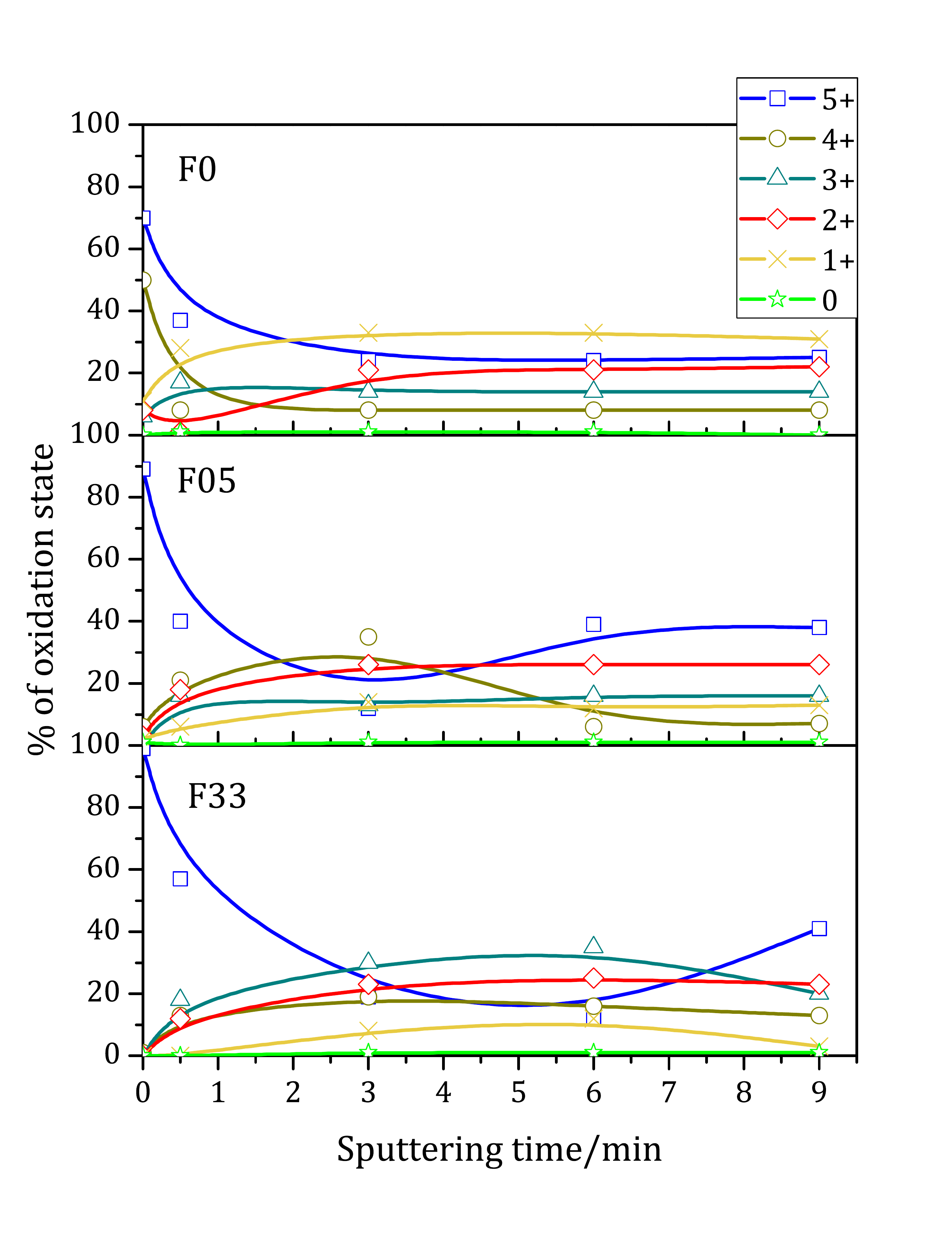}
		\caption{Percentage of oxidation state of Ta as a function of sputtering time for the three films. Lines are just a guide to the eye}
		\label{percentage}
	\end{center}
\end{figure}

To quantify the contribution of Ta oxide to the whole spectrum, we computed, from the deconvolution of the Ta $4f$ spectra, the percentage of oxidation state of Ta (see Figure \ref{percentage}). 
Due to the complexity of the spectra, the peak fitting may not be entirely accurate and over interpretation of the results should be avoided. Nonetheless, some general trends can be indeed summarized. In Figure \ref{percentage} we can observe that, at the surface of all samples, Ta in Ta$^{5+}$ has the major contribution; with more than 70\% in all cases. As expected the content of Ta$^{5+}$ in F0 is much less than that one in F05 or F33. As the irradiation time increases, the state +5 decreases between 15\% and 35\% whereas the other states increase, suggesting the presence of Ta suboxides. Ta in metallic state (Ta$^{0}$) is barely present in all cases with less than 1\% and the presence of suboxides varies from 10\% to 35\%. These values are reasonable according to previous reports for amorphous samples \cite{rsimpson17a}. 

\begin{figure}[t!]
	\begin{center}
		\includegraphics[width=11cm]{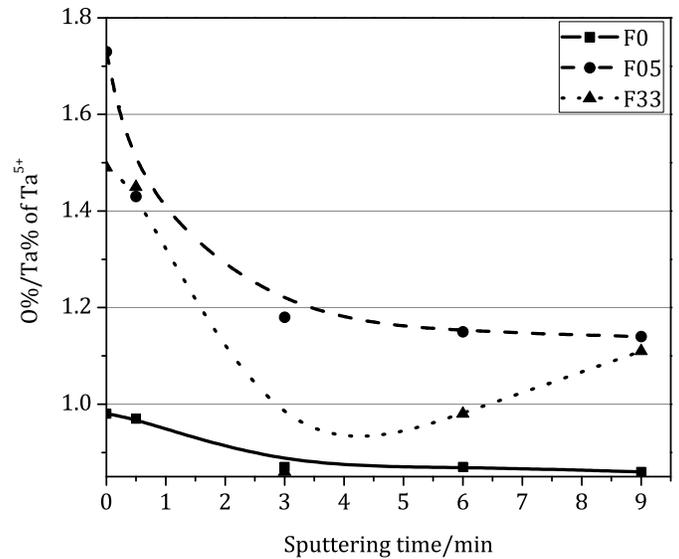}
		\caption{O at\% to Ta at\% ratio for Ta$^{5+}$ (Ta$_2$O$_5$) as a function of sputtering time for the three coatings. Lines are just a guide to the eye}
		\label{ratio}
	\end{center}
\end{figure}

We also estimated the stoichiometry of our samples using the typical expression for composition $C_x$ of an atomic species $x$:
\begin{equation}
\label{compo}
C_x=\frac{\frac{I_x}{SF_x}}{\sum_{i=1}^N\frac{I_i}{SF_i}},
\end{equation}
where $I_i$ are the $i$ peak intensities for each atomic species, that is, Ta and O; and $SF_i$ is the sensitivity factor for the atomic species $i$. The sensitivity factors are given by the XPS instrument ( $SF_O=2.93$ and $SF_{\text{Ta}}=8.62$) and the peak intensities were obtained during the deconvolution process. In stoichiometric Ta$_2$O$_5$ the O to Ta ratio is 2.5. We have plotted the atomic percentage ratio for Ta$^{5+}$ in figure \ref{ratio}. There we can see that all samples are non-stoichiometric. The ratio of F0 ranges from 0.80 to 0.95 as sputtering time increases which seems reasonable since this sample was not exposed to a heat treatment and one would expect the same atomic concentration all over the sample (except at the surface where there is slightly more oxygen available). 
This finding suggests that the composition of this film is mainly TaO$_x$, reaffirming our discussion above. The ratio for F05 goes from 1.50 to 0.90 as the sputtering time increases and that one for F33 goes from 1.75 to 1.20; these values indicate an overall low oxygen content. The non-stoichiometry is likely to be consequence of two important factors, namely: the lack of an oxygen rich atmosphere during the manufacturing process of the samples; and the oxygen depletion taking place during ion bombardment. It is evident from the plot that the latter factor has a major impact on the oxygen depletion. 

We remark that, according to our analysis, most Ta suboxides only appear after sputtering the film surface (even for sputtering times as short as 30 s). If they were present all over the bulk of the film, they would show up in the XRD patterns; however the XRD patterns show no signs of impurity phases of Ta oxide (of course, excluding Ta$_2$O$_3$ and TaO$_x$ which are amorphous and thus cannot be detected by XRD). To further check these assertions, after sputtering F33 for 9 min, we exposed it to air for three days and obtained again its Ta $4f$ spectrum at the surface; the results showed anew the spectrum corresponding to amorphous Ta$_2$O$_5$. This spectrum is quite similar to the one displayed in Figure \ref{ta4fo1s} and therefore it is not shown here.

\subsubsection{O $1s$ core level}
We can extract even more information on the chemical states analyzing the oxygen core level. The results for the O $1s$ core level for all films are displayed in the right column of Figure \ref{ta4fo1s}. Starting with the black spectra (no etching) we observe a hump at high energies, around ($533.8 \pm0.2$) eV. We also note a prominent signal at 532 eV which is typical from Ta-O binding \cite{amuto94a,hszymanowski05a}. 
\begin{figure*}[t!]
	\begin{center}
		\includegraphics[width=17.4cm]{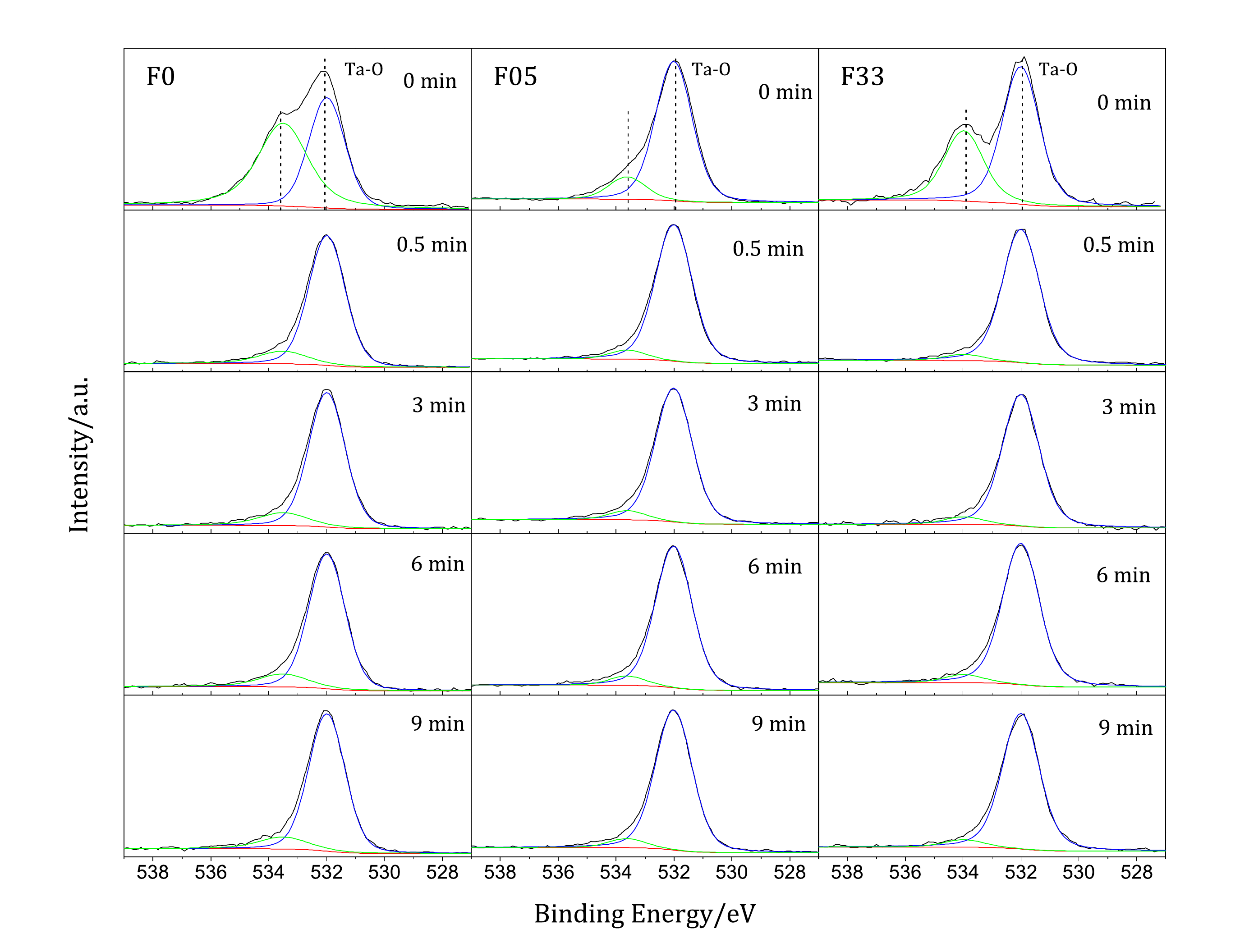} 
		\caption{Deconvolution of O $1s$ XPS spectra for the films. Left column for F0, central column for F05, and right column for F33}
		\label{deconvo1s}
	\end{center}
\end{figure*}
The situation for the sputtered films is different for the hump seems to vanish. The deconvolution unveils the presence of two components, indicating that even for bottom layers this signal is still present although its intensity is diminished (see figure \ref{deconvo1s}). This feature suggests C-O binding and has been observed before by some researchers \cite{eatanassova98a,ttsuchiya11a,okerrec98a}. On one hand, O. Kerrec et al. grew amorphous Ta$_2$O$_5$ films and exposed them to distilled water. They attributed the feature to hydroxylic groups and/or water adsorbed at the surface. On the other hand, Atanassova et al. grew also amorphous Ta$_2$O$_5$ thin films (6 nm to 13 nm) on Si substrates. They explained that the feature was due to Si-O binding rather than contamination. In a more recent investigation, R. Simpson et al. \cite{rsimpson17a} studied the effect on the chemical composition of Ar$^+$ sputtered amorphous Ta$_2$O$_5$ films (30 nm thick) grown on Ta foil and found that the C $1s$ core level disappears after 3 nm depth. Thus since our coatings are thick and the intensity of the feature diminishes considerably as the sputtering time increases, we attribute it to surface contamination, must probably a carbon compound (CO) due to the presence of carbon on the surface. We rule out the formation of Ta-C binding because is well known that tantalum carbide crystallizes at temperatures larger than 1600$^\circ$C \cite{grgruzalski86a}.

\section{Conclusions}
The results of SEM showed a granular structure for the crystalline samples resembling the powder morphology. Using XRD, the samples were indexed to the orthorhombic phase of Ta$_2$O$_5$ for the crystalline films and, accordingly, we observed no traces of crystalline Ta suboxides in the bulk of all samples. Meanwhile F0 was found to be amorphous. In this sense, the crystalline analysis indicates, after the heat treatment, a structural transition of F05 and F33 from Ta amorphous to crystalline Ta$_2$O$_5$. We carried out XPS analysis in order to study the effect of ion irradiation on the chemical properties of our samples for several time intervals. From the deconvolution of the Ta $4f$ spectra it can be concluded that the chemical states of Ta vary from Ta$^{5+}$ to Ta$^{1+}$ with a nil contribution of the oxidation state Ta$^{0}$ when the samples are irradiated. There is an overall reduction of the oxygen to tantalum ratio as the sputtering time increases which is caused by the high oxygen sputtering yield. Our findings strongly suggest the manifestation of Ta suboxides generated at the surface derived by the creation of Frenkel defects and atom displacement mainly during ion bombardment as indicated by the simulation. We found evidence that the suboxides are unstable and the surface composition can be restored to the amorphous phase of Ta$_2$O$_5$ after several days of air exposure. We conclude that all films, after the ion bombardment, exhibit an amorphous Ta$_2$O$_5$ phase at the surface although some degree of non-stoichiometry is present. We mainly attribute this to oxygen depletion during irradiation and to the lack of an oxygen rich environment during both growth and annealing. We have demonstrated that ion irradiation induces the formation of several unstable oxidation states of tantalum.

\section*{Acknowledgements}
We are grateful to Wilian Cauich, Daniel Aguilar, and Jorge Ivan Betance for their technical support during the XPS, XRD, and SEM sessions. We are grateful to the editor of this journal and the anonymous reviewers for raising several comments that greatly improved the quality of this research. Prof. I. Perez is thankful to Dr. Alberto Herrera for helpful comments and discussions on the XPS analysis. Funding: This work was partially supported by the National Council of Science and Technology (CONACYT) Mexico and the program C\'atedras CONACYT through project 3035.

{\bf Declaration of interest:} none

\end{document}